\documentstyle[12pt,psfig,epsf]{article}
\begin{document}
\thispagestyle{empty}
\newcommand{\ima}{\hbox{Im}\;}
\newcommand{\rea}{\hbox{Re}\;}
\centerline{\hfill NUHEP--TH--92--17}
\pagestyle{empty}
\begin{center}
{\bf CP Violation in top pair production\\
at an $e^+e^-$ collider}
\end{center}
\vskip 1cm
\centerline{
Darwin Chang$^{(1,2)}$,
Wai--Yee Keung$^{(3,4)}$,
Ivan Phillips$^{(1)}$}
\vskip 1cm
\begin{center}
$^{(1)}$Department of Physics and Astronomy,\\
Northwestern University, Evanston, IL 60208, USA.\\
$^{(2)}$Institute of Physics, Academia Sinica, Taipei, Taiwan, R.O.C.\\
$^{(3)}$Physics Department, University of Illinois at Chicago,
IL 60680, USA \\
$^{(4)}$Theory Division, CERN, CH-1211, Geneva 23, Switzerland \\
\end{center}
\vskip 1cm
\begin{abstract}
We investigate possible CP violating effects in $e^+e^-$
annihilation into top quark pairs.
One of the interesting observable effects is the difference in production
rates between the two CP conjugate polarized states $t_L\bar t_L$
and $t_R\bar t_R$. The result is an asymmetry in the energy spectra of
the lepton and the anti--lepton from the heavy quark decays.
Another CP--odd observable is the up-down asymmetry of the leptons with
respect to the reaction plane.
%
%
These two asymmetries measure complementarily the absorptive and
dispersive form factors of the electric dipole moment.
Finally, as an illustration, we calculate the size of the CP violating form factors in a
model where the CP nonconservation originates from the Yukawa couplings
of a neutral Higgs boson.
\end{abstract}
\vskip 2cm

\noindent
Published in Nucl. Phys. {\bf B408} (1993) 286; 
Erratum, {\it ibid.}, {\bf B429} (1994) 255.

%
\newpage
%
%
\pagestyle{plain}

\noindent
{\bf I. Introduction.}
 
Since the top quark is widely believed to be within the reach of the present
collider machines, it is not unreasonable for theorists to imagine what we
can learn from the top quark.  The best place to study the top quark in detail
is in an $e^+e^-$ collider.  One of the facts one would like to learn from the
discovery of the top quark is the origin of the still mysterious CP violation.
In this paper we investigate a way CP violation can manifest itself in the
top pair production of an $e^+e^-$ collider.
 
The top quark, due to its short lifetime, is believed to decay
before it hadronizes\cite{ref:Bigi}. Therefore the information about its
polarization may be preserved in its decay products.  If that is the
case, then one can investigate the source of CP nonconservation by
measuring the CP violating observable involving a polarized top pair in
the final state.  This idea of detecting the rate asymmetry between
different polarized states was recently proposed by Schmidt and 
Peskin\cite{ref:Peskin,ref:DK}.
For $t {\bar t}$ production through the virtual photonic or $Z$
intermediate states, to the lowest order in the final state quark mass,
the polarizations of the quarks are either $t_L {\bar t}_R$ or $t_R
{\bar t}_L$. (Note that we have adopted the notation that ${\bar t}_L$
is the antiparticle of $t_R$ and should be left handed.) These two modes
are CP self-conjugate.  However since the top quark is heavy, there
will also be large percentage of $t_L {\bar t}_L$ and $t_R {\bar t}_R$
modes which are CP conjugates of each other. Therefore one can consider
a CP asymmetry in the event rate , $N(t_L {\bar t}_L)-N(t_R
{\bar t}_R)$.
 
Schmidt and Peskin have shown how to detect the asymmetry
$N(t_L {\bar t}_L)-N(t_R {\bar t}_R)$
through the energy spectra of prompt leptons\cite{ref:Peskin,ref:DK}.
One assumes that
the $t$ quark decays semileptonically through the usual $V-A$ weak
interaction.
Knowing that the hadronization time is much longer than the decay
time\cite{ref:Bigi}, one can analyze polarization dependence of its
decay at the quark level. The top quark first decays into a $b$ quark
and a $W^+$ boson, which subsequently becomes $\ell^+ \nu$.
For heavy top quark, the $W^+$ boson produced in top decay is predominantly
longitudinal.
Due to the $V-A$ interaction, the $b$ quark is preferentially produced
with left-handed helicity.  So the longitudinal $W^+$ boson is
preferentially produced along the direction of the top quark polarization.
Therefore the anti--lepton $\ell^+$ produced in the $W^+$ decay is also
preferentially in that direction.
In the rest
frame of the $t$, the angular distribution\cite{ref:Kuhn}
of the produced $\ell^+$ has the
form $1+\cos\psi$, with $\psi$ as the angle between $\ell^+$ and the
helicity axis of the $t$.
Above the $t {\bar t}$ threshold, the top quark is produced with nonzero
momentum. As a result of the Lorentz boost, the anti--lepton
$\ell^+$ produced in the decay of the right handed top quark $t_R$ has
a higher energy than that produced in the decay of the left handed top
quark $t_L$. Similarly, the lepton $l$ produced in the decay of ${\bar
t}_L$ has a higher energy than that produced in the decay of  ${\bar
t}_R$. Consequently, in the decay of the pair $t_L {\bar t}_L$ the
lepton from $\bar t_L$ has a higher energy than the anti--lepton
from $t_L$; while in the decay of $t_R {\bar t}_R$ the anti--lepton
has a higher energy.
Therefore one can observe $N(t_L {\bar t}_L)-N(t_R {\bar t}_R)$
by measuring the energy asymmetry in the
leptons. It turns out that this energy asymmetry is sensitive only to
the absorptive parts of CP violating form factors.
 
There is another equally interesting CP--odd effect in the azimuthal angular
distribution, namely, the rate difference between the events with $\ell^\pm$
above the reaction plane and the events with $\ell^\pm$ below the reaction
plane. Such an observable, like the previous one, is a direct measurement
of CP violation and thus has no background from the CP conserving interactions.
Unlike the previous case, this up--down asymmetry will probe the CP violating
dispersive form factors.   
%
%
Though there have been many studies of CP violating observables in the
literature\cite{ref:OtherCP,ref:KLY}, we feel that the above two
observables are simple and intuitive in nature and are easily
implemented in future experiments. 
 
Finally, we will use a generic neutral Higgs model with CP violation
in the Yukawa couplings to illustrate how these observables can arise.
Such mechanisms of CP violation are contained in many extensions of
Standard Model including the simple two doublet model\cite{ref:doublet}.
Throughout this paper, we focus our attention on CP
non--conservation in the production mechanism only.
There will be additional contributions if the usual $V-A$ decay amplitude
is also modified by CP violating interactions\cite{ref:decayCP}.
 
\noindent
{\bf II. CP violating form factors and amplitudes.}
 
   We start by writing down the general form factors of the $t$ quark.
The vertex amplitude $ ie \Gamma^j$ for the virtual
$\gamma^*$ or $Z^*$ turning into $t(p)$ and $\bar t(p')$ can be
parametrized in the following expression:
\begin{equation}
  \Gamma^j_\mu  = c_v^j \gamma_\mu  + c_a^j \gamma_\mu\gamma_5
                + c_d^j i\gamma_5 {p_\mu-p'_\mu\over 2 m_t} + \cdots,
\quad j=\gamma,Z.
\label{eqn:form}
\end{equation}
We use the tree--level values for  $c_v$ and $c_a$. They are
\begin{eqnarray}
 c_v^\gamma   &=&\hbox{$2\over3$}, \quad\quad c_a^\gamma=0,      \nonumber\\
 c_v^Z  &=&(\hbox{$1\over4$} - \hbox{$ 2\over3$} x_W)/\sqrt{x_W(1-x_W)}\;,\\
 c_a^Z  &=& -\hbox{$1\over4$}           /\sqrt{x_W(1-x_W)}
\; .  \nonumber
\end{eqnarray}
Here $x_W\simeq 0.23$ is the electroweak mixing angle in the Standard
Model.
The $c_d$ terms are the electric dipole form factors. The spinor
structure can be rewritten into another form using
$i\gamma_5(p-p')_\mu = \sigma_{\mu\nu}(p+p')^\nu\gamma_5$.
Other irrelevant terms, like the magnetic moments, are not listed in
Eq.(\ref{eqn:form}).
It can also be easily shown that,
in the limit $m_e=0$, $c_d$ is the only relevant form factor for the CP
violating quantities we are interested in.

The helicity amplitudes $e^2 M(h_e,h_{\bar e},h_t,h_{\bar t})$
for the process $e^-e^+\rightarrow t \bar t$ at the scattering
angle $\theta$ have been given in the literature\cite{ref:KLY}.
For the initial configuration of $e_L\bar e_R$, we have
\begin{eqnarray}
M(-+-+)=&[c_v^\gamma+r_L c_v^Z-\beta r_L c_a^Z](1+\cos\theta) \nonumber\\
M(-++-)=&[c_v^\gamma+r_L c_v^Z+\beta r_L c_a^Z](1-\cos\theta) \nonumber\\
M(-+--)=&[2t(c_v^\gamma+r_L c_v^Z)
           -\hbox{$ i\over2$} (c_d^\gamma + r_L c_d^Z)\beta /t] \sin\theta
\label{eq:amp}
                                                                    \\
M(-+++)=&[2t(c_v^\gamma+r_L c_v^Z)
           +\hbox{$ i\over2$} (c_d^\gamma + r_L c_d^Z)\beta /t] \sin\theta
                                                                 \;.\nonumber
\end{eqnarray}
Here we have used the convention \cite{ref:phase}
that CP invariance, when $c_d^{\gamma,Z}$ are turned off,
is signified by the relation
\begin{equation}
M(\sigma,\bar{\sigma};\lambda,\bar{\lambda})=
M(-\bar{\sigma},-\sigma;-\bar{\lambda},-\lambda).
\label{eq:CP}
\end{equation}
The dimensionless variables are defined by, $t=m_t/\sqrt{s}$,
$z=m_Z/\sqrt{s}$, $\beta^2=1-4t^2$. The $Z$--propagator and its
coupling to the left--handed electron gives $-e r_L/s$ with
\begin{equation}
      r_L=(\hbox{$ 1\over2$} - x_W)/ [(1-z^2) \sqrt{x_W (1-x_W)}] \;
\end{equation}
The cross section is
\begin{equation}
       d  \sigma(e_L \bar e_R \rightarrow t_{h_t} \bar t_{h_{\bar t}})
     / d(\cos\theta)
=
     \hbox{$ 3\over2$} \pi\alpha^2\beta |M(-+,h_t,h_{\bar t})|^2 /s \;.
\end{equation}
A color factor $3$ has been included explicity in the above formula.
Similarly, we obtain formulas for the initial configuration $e_R\bar e_L$
with $r_L$ replaced by $r_R$,
\begin{equation}
      r_R=-x_W /[(1-z^2) \sqrt{x_W (1-x_W)}] \; ,
\end{equation}
and $\cos\theta$ by $-\cos\theta$, and $\sin\theta$ by $-\sin\theta$ 
in Eq.(\ref{eq:amp}).
In case of an unpolarized $e^-e^+$ machine, we must
sum up initial configurations $e_L\bar e_R$ and $e_R\bar e_L$, and
include the spin average factor $1\over 4$.
 
It is also interesting to note that, if the absorptive part of a scattering
amplitude can be ignored (which is certainly true at the tree level),
then the unitarity of the S matrix implies that
the scattering matrix is hermitian.  This hermiticity allows one to write
down the constraint due to the CPT invariance as
\begin{equation}
M(\sigma,\bar{\sigma};\lambda,\bar{\lambda})=
M^*(-\bar{\sigma},-\sigma;-\bar{\lambda},-\lambda).
\label{eq:CPT}
\end{equation}
We shall refered to this special case of CPT invariance as CP$\hat{T}$
invariance.  The CP$\hat{T}$ invariance of course can be violated by the
absorptive part of the scattering amplitude.
 
\noindent
{\bf III. Leptonic Energy asymmetry}
 
It is straightforward to see that the absorptive part
$\ima c_d^\gamma$ or $\ima c_d^Z$ is required to produce
the difference between configurations $t_L\bar t_L$ and $t_R\bar t_R$.
This is expected because we need the absorptive part
via the final state interactions to overcome the CP$\hat{T}$ constraint.
The asymmetry integrated over the angular distribution is
\begin{equation}
  \delta \equiv {[N(t_L\bar t_L) -N(t_R\bar t_R)]
    \over N(t\bar t; \hbox{all}) }
= { \sum_{h=L,R} 2\beta(c_v^\gamma + r_h c_v^Z)
                       (\ima c_d^\gamma + r_h \ima c_d^Z)
            \over
\sum_{h=L,R} (3-\beta^2)(c_v^\gamma + r_h c_v^Z)^2 +
               2\beta^2 r_h^2  {c_a^Z}^2}
\;.
\label{eq:asymE}
\end{equation}
Note that there is no CP violating contribution
due to $c_a^Z$ coupling in the numerator when the electron mass $m_e$
is ignored.

We can make use of this asymmetry parameter $\delta$ to illustrate the
the difference in the energy distributions of $\ell^+$ or $\ell^-$
from the $t$ or $\bar t$ decays.
The energy $E_0(\ell^+)$ distribution of a static $t$ quark decay
$t\rightarrow \ell^+\nu b$ is very simple\cite{ref:Kuhn}
in the narrow width $\Gamma_W$
approximation when $m_b$ is negligible.
It can be represented as
\begin{equation}
    f(x_0)= \left\{
              \begin{array}{lll}
       x_0 (1-x_0)/D & \quad\  & \mbox{if $m_W^2/m_t^2 \leq x_0 \leq 1$}, \\
       0             & \quad\  & \mbox{otherwise.}
             \end{array}
              \right.
\end{equation}
for on-shell W in the decay.
Here we denote the scaling variable $x_0=2E_0(\ell^+)/m_t$ and
the normalization factor $D={1\over 6}-{1\over 2}(m_W/m_t)^4
+{1\over 3}(m_W/m_t)^6$.
When the $t$ quark is not static, but moves at a speed $\beta$
with helicity $L$ or $R$, the distribution expression becomes
a convolution,
\begin{equation}
     f_{R,L}(x,\beta)=
    \int_{x/(1+\beta)}^{x/(1-\beta)} f(x_0)
            {\beta x_0 \pm (x-x_0) \over 2 x_0^2\beta^2}
      dx_0
\;.
\end{equation}
Here $x=2E(\ell^+)/E_t$. The kernel above is related to the
$(1\pm\cos\psi)$  distribution mentioned in the introduction.
Similar distributions for the $\bar t$ decay are related by CP
conjugation at the tree--level.
Using the polarization asymmetry formula in  Eq.(\ref{eq:asymE}),
we can derive an expression for the difference in the energy
distributions of $\ell^-$ and $\ell^+$:
\begin{equation}
{1\over N} \Bigl[ {dN\over dx(\ell^+)}-{dN\over dx(\ell^-)} \Bigr]
=\delta[f_L(x,\beta)-f_R(x,\beta)]
\;.
\label{eq:disd}
\end{equation}
Here distributions are compared at the same energy for the lepton
and the anti--lepton, $x(\ell^-)=x(\ell^+)=x=4E(\ell^\pm)/\sqrt{s}$.
The count $N$ includes events with
prompt leptons or anti--leptons from the top pair production.
It is useful to compare
Eq.(\ref{eq:disd}) with that of the overall energy distribution,
$$
{1\over N} \Bigl[ {dN\over dx(\ell^+)}+{dN\over dx(\ell^-)} \Bigr]
={  \sum_{h=L,R} 4\beta r_h c_a^Z(c_v^\gamma +r_h c_v^Z)
                                       [f_R(x,\beta)-f_L(x,\beta)]
                        \over
\sum_{h=L,R} (3-\beta^2)(c_v^\gamma + r_h c_v^Z)^2 +
               2\beta^2 r_h^2  {c_a^Z}^2}
$$
\begin{equation}
 \quad\quad
           +  f_L(x,\beta) + f_R(x,\beta)
\;.
\label{eq:diss}
\end{equation}
Here we only keep the dominant tree--level contribution.
The first term of Eq.(\ref{eq:diss}) is due to the two helicity modes
that are CP self-conjugate.

\centerline{\psfig{figure=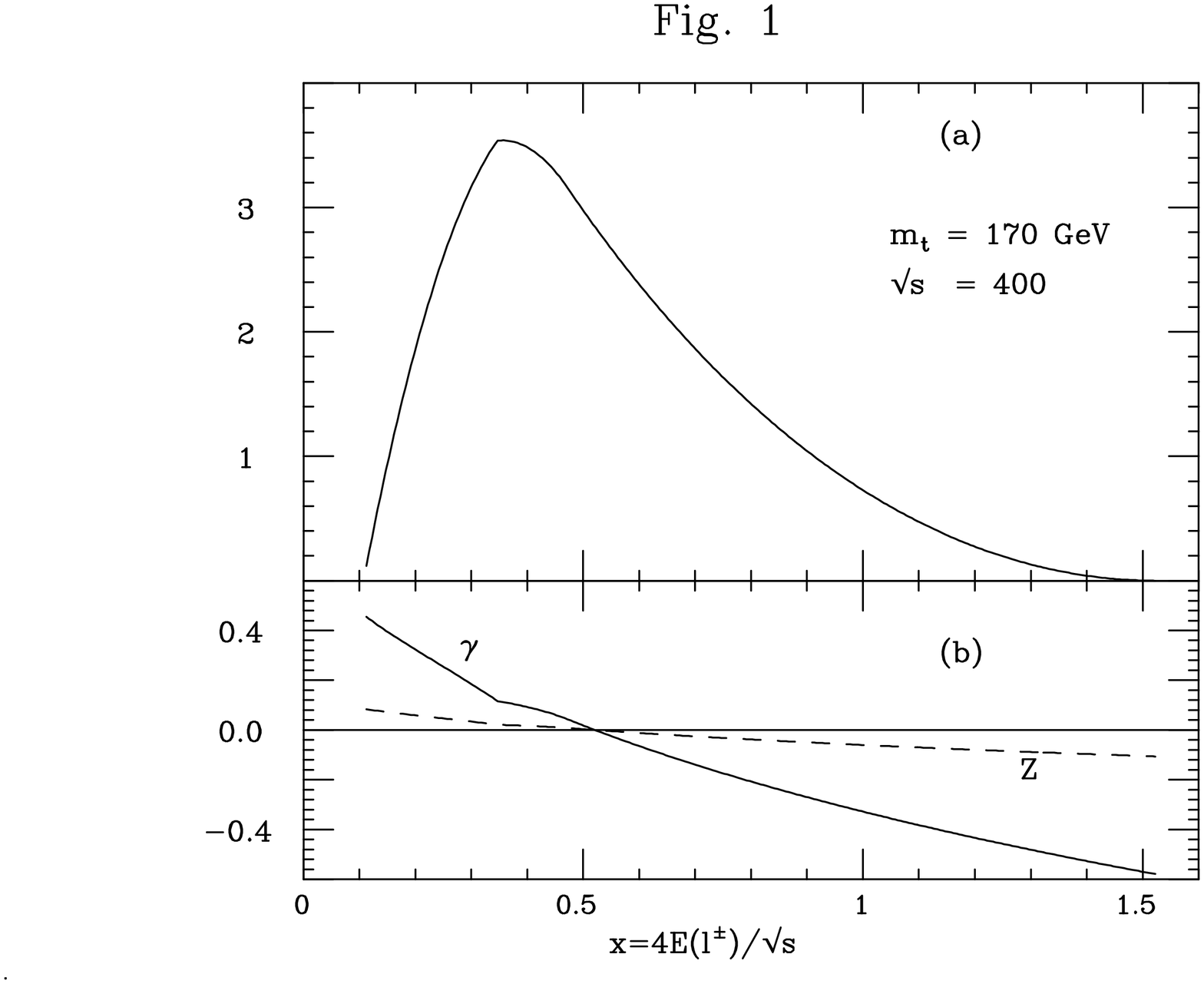,height=4in,width=6.5in,angle=90}}
Fig.~1 shows the overall prompt lepton energy distribution
of Eq.(\ref{eq:diss}),
and the ratio of the expressions in Eq.(\ref{eq:disd}) and
Eq.(\ref{eq:diss}) per unit $\ima c_d^\gamma$ or
$\ima c_d^Z$.
 
Note that our sample includes events with one or at least one prompt
$\ell^\pm$.   
However, if we are willing to use a smaller sample of prompt dilepton
events (from the simultaneous semileptonic 
decays of both $t$ and $\bar t$), a recent paper 
Ref.~\cite{ref:LY} has shown how one can 
use the kinematics of such events to analyze the helicities of 
$t$ and $\bar t$.  Therefore such knowledge can be used to 
enhance the detectability of the asymmetry by
cutting away the CP self-conjugate modes in the denominator of
$\delta$ in Eq.(\ref{eq:asymE}) \cite{ref:LY}.  However, the reduction 
in the size of event samples may be too high a price to pay.  

\noindent
{\bf IV. CP--odd up--down asymmetry.}
 
It is known that explicit CP violation requires the CP nonconserving
vertex as well as additional complex amplitudes. In the above case, this
complex structure comes from the absorptive part due to the final state
interactions. However, the complex structure can also come from other
sources. One of these is the azimuthal phase $\exp(iL_z\phi)$ in the
decay process. This will produce a CP--odd up--down
asymmetry even with only the dispersive part of CP violating vertex,
$\rea c_d$.
 
The angular distribution of $\ell^+$ from the $t$ decay is specified by the
the spin density matrix $\rho_{\lambda,\lambda'}$ of the top quark.
\begin{equation}
   \rho(\theta)_{\lambda,\lambda'}
      ={\cal N}(\theta)^{-1}\sum_{h_e,h_{\bar e},h_{\bar t}}
    M(h_e,h_{\bar e},\lambda ,h_{\bar t})
  M^*(h_e,h_{\bar e},\lambda',h_{\bar t})\;.
\end{equation}
Here ${\cal N}$ is the normalization such that $\hbox{Tr}\rho$=1.
$\rho$ is hermitian by definition.
\begin{equation}
   dN(\ell^+) =\Bigl[(1+\cos\psi)\rho_{++}
                + (1-\cos\psi)\rho_{--}
                +2\rea (\rho_{+-}e^{i\varphi})\sin\psi \Bigr]
                  d\varphi d\cos\psi/(4\pi)    \;.
\label{eq:phi}
\end{equation}
The polar angles $\psi$ and the azimuthal angle $\varphi$ of $\ell^+$
are defined in the $t$ rest frame $F_t$, which is constructed by
boosting the $t\bar t$ center of mass frame $F_0$ along the
motion of the top quark. In $F_0$, the $z$--axis is along the top
momentum ${\bf p}_t$, the $y$ axis is along
${\vec p}_{e^-} \times {\vec p}_t$,
and the $x$-axis is given by the right--handed rule.
The production plane is the $x-z$ plane.
 
Similarly, the angular distribution of $\ell^-$ from the $\bar t$ decay is
specified by the spin density matrix,
$\bar\rho_{\bar \lambda,\bar\lambda'}$, of the anti--top quark.
\begin{equation}
   \bar \rho(\theta)_{\bar \lambda,\bar \lambda'}
      ={\cal N}(\theta)^{-1} \sum_{h_e,h_{\bar e},h_t}
    M(h_e,h_{\bar e},h_t,\bar \lambda)
  M^*(h_e,h_{\bar e},h_t,\bar \lambda')\;.
\end{equation}
\begin{equation}
   dN(\ell^-) =\Bigl[(1+\cos\bar\psi)\bar\rho_{++}
                + (1-\cos\bar\psi)\bar\rho_{--}
                 -2\rea (\bar\rho_{+-}e^{-i\bar\varphi})\sin\bar\psi \Bigr]
                  d\bar\varphi d\cos\bar\psi/(4\pi)    \;.
\label{eq:phibar}
\end{equation}
The polar angles $\bar \psi$ and the azimuthal angle $\bar \varphi$
of $\ell^-$ are defined in the $\bar t$ rest frame $F_{\bar t}$,
which is similarly constructed by
boosting the $t\bar t$  center of mass frame $F_0$ along the
motion of the anti--top quark. It is important to keep in mind that
the three coordinate systems $F_0$, $F_t$ and $F_{\bar t}$,
have parallel directions of coordinate axes.
 
In the phase convention\cite{ref:phase}
such that Eqs.(\ref{eq:CP},\ref{eq:CPT}) are satisfied when CP is
conserved,
the following identities, first noticed by
Gounaris et al.\cite{ref:Gounaris},
\begin{equation}
\rho(\theta)_{\lambda,\lambda'}=\bar\rho(\theta)_{-\lambda,-\lambda'}
\; ,
\label{eq:Gou}
\end{equation}
between the density matrix elements can be derived.
Under CP conjugation, as we exchange $\ell^-$ and $\ell^+$,
by definition one should
make the angular substitutions $\bar\psi \rightarrow \pi-\psi$,
$\bar\varphi \rightarrow \pi+\varphi$.
In that case, one observes that the distribution in Eq.(\ref{eq:phibar})
is transformed into Eq.(\ref{eq:phi}) provided Eq.(\ref{eq:Gou}) is
satisfied.
On the other hand, the CP$\hat{T}$ invariance, when the absorptive
amplitudes are ignored,  implies
\begin{equation}
\rho(\theta)_{\lambda,\lambda'}=\bar\rho(\theta)^*_{-\lambda,-\lambda'}
\; ,
\label{eq:Goucpt}
\end{equation}
This is very similar to the
CP and CP$\hat{T}$ transformations in the case of $e^-e^+\rightarrow W^-W^+$
as analyzed before\cite{ref:Gounaris,ref:Hagiwara}.

When the effect of the CP violating form factors are included in the
analysis, one can form the following CP or CP$\hat{T}$ odd combinations:
$$R(\pm)(\theta)_{\lambda,\lambda'} =
\rea \rho(\theta)_{\lambda,\lambda'}
\pm \rea \bar\rho(\theta)_{-\lambda,-\lambda'}   \;,$$
and
$$I(\pm)(\theta)_{\lambda,\lambda'} =
\ima \rho(\theta)_{\lambda,\lambda'}\pm
\ima \bar\rho(\theta)_{-\lambda,-\lambda'}       \;.$$
Among them, $R(-)$ and $I(+)$ are CP$\hat{T}$ odd;
$R(-)$ and $I(-)$ are CP odd and all the others CP and CP$\hat T$ even.
The observation of $R(-)$ requires final state interactions
due to CP$\hat T$.  Therefore it does not need to involve the the complex
phase of the azimuthal dependence in Eqs.(\ref{eq:phi}, \ref{eq:phibar}).
It can be decoded by analyzing the polar angular dependence
of $\psi$ or $\bar\psi$ in these equations.
In the collider C. M. frame these
dependence can be translated into the
energy dependence of the corresponding lepton in the final state,
which has already been studied in the previous section.
 
 
Here we will focus on $I(-)$ which does not require
final state interactions.  Since $\rho$ is hermitian, the only nonzero
component of $I(-)$ is $I(-)_{+-}$.  It can be related to
$c_d$ as
\begin{equation}
{\cal N}(\theta) \ima [\rho(\theta)_{+-} - \bar\rho(\theta)_{-+}]
= 2  \sin\theta\sum_{h=R,L}(-h) (c_v^\gamma+r_h c_v^Z \pm
                                 \beta r_h c_a^Z\cos\theta)
             \rea (c_d^\gamma+r_h c_d^Z)\beta/t \;,
\label{eq:difden}
\end{equation}
where the contributions $h=R,L$ pick up the signs $+,-$ respectively.
One can in principle make detailed angular analysis of the difference
between Eq.(\ref{eq:phi}) and its CP conjugate in Eq.(\ref{eq:phibar})
similar to what was done for the case of $e^-e^+\rightarrow W^-W^+$
by Gounaris et al.\cite{ref:Gounaris}
However, in an effort to find simpler observables which may be more
intuitive and may be easier to detect, we shall consider the following
partially integrated observable.
Let $dN(\ell^+,\hbox{up})$ count events with $\ell^+$ above the $xz$ plane,
{\it i.e.} $p_y(\ell^+)>0$. Then,
with other obvious notations, we define the following up--down asymmetry
\begin{equation}
{\cal A}^{u.d.}(\theta) =
{
 [dN(\ell^+,\hbox{up})+dN(\ell^-,\hbox{up})]
-[dN(\ell^+,\hbox{down})+dN(\ell^-,\hbox{down})]
\over
 [dN(\ell^+,\hbox{up})+dN(\ell^-,\hbox{up})]
+[dN(\ell^+,\hbox{down})+dN(\ell^-,\hbox{down})]
}\;.
\label{eq:Aud}
\end{equation}
It is evaluated for each scattering angle $\theta$. The branching
fraction of the $t$ semileptonic decay cancels in the ratio. Integrating
on $\psi,\varphi$ or $\bar\psi,\bar\varphi$ over up or down hemispheres,
we obtain ${\cal A}^{u.d.}(\theta)$ in a very simple form from
Eqs.(\ref{eq:phi},\ref{eq:phibar}),
\begin{equation}
{\cal A}^{u.d.}(\theta)=\hbox{$1\over2$}(
   \ima \bar\rho(\theta)_{-+}-\ima \rho(\theta)_{+-})
\label{eq:Ad}
\end{equation}
As we sum up contributions from $\ell^\pm$ in each square bracket of
Eq.(\ref{eq:Aud}),
the asymmety is insensitive to the sign of charge,
it is obvious that a non-vanishing value
of ${\cal A}^{u.d}(\theta)$ is a genuine signal
of CP violation. Although the angular distributions of the leptons
derived from Eq.(\ref{eq:amp}) will have corrections from the strong
interaction, the corrections {\it cannot} fake the CP asymmetry as the
effects due to the strong interaction cancel away in the differences.

To enhance statistics, it is useful to measure the integrated up--down
asymmetry,
\begin{equation}
{\cal A}^{u.d.}={1\over \sigma}
\int {\cal A}^{u.d.}(\theta) {d\sigma\over d(\cos\theta)} d(\cos\theta)
\; .
\label{eq:Ai}
\end{equation}
From Eqs.(\ref{eq:difden}-\ref{eq:Ai}), we obtain,
\begin{equation}
{\cal A}^{u.d.}=-{3\pi\over 16}
 { \sum_{h=L,R} (-h) (c_v^\gamma + r_h c_v^Z)
                       (\rea c_d^\gamma + r_h \rea c_d^Z)\beta/t
            \over
\sum_{h=L,R} (3-\beta^2)(c_v^\gamma + r_h c_v^Z)^2 +
               2\beta^2 r_h^2  {c_a^Z}^2}
\;.
\label{eq:asymA}
\end{equation}
Note that information on $c_a^Z$ in Eq.(\ref{eq:difden}) is lost when
integrating the whole range of $\cos\theta$. However, one can easily
find other convolution in Eq.(\ref{eq:Ai}) to recover the CP information
due to $c_a^Z$.
 
\centerline{\psfig{figure=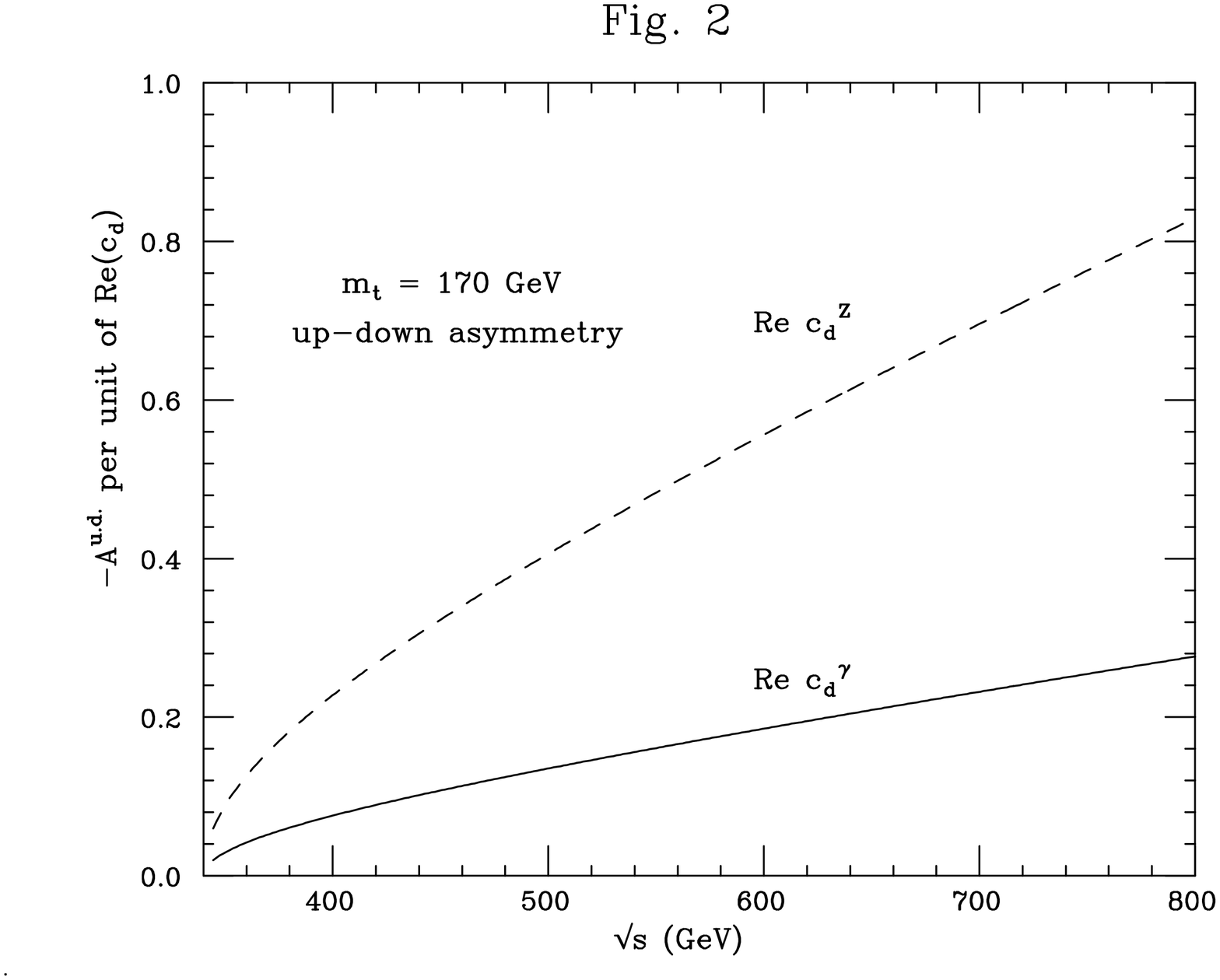,height=4in,width=6.5in,angle=90}}
In Fig.~2, we plot the integrated up--down asymmetry ${\cal A}^{u.d.}$
per unit $\rea c_d^\gamma$ or $\rea c_d^Z$ for various parameters.
It seems that ${\cal A}^{u.d.}$ per unit $\rea c_d$ increases
with $\sqrt{s}$, however, Re $c_d$ is usually energy dependent,
characterized by the underlying scale of new physics. Above that scale,
Re $c_d$ diminishes very fast. Therefore, the optimal choice of
$\sqrt{s}$ is about at the scale of new physics. Note that even if
Re $c_d$ is relatively constant and the angular asymmetry is 
larger at higher energy, the event rate will become smaller 
because of the nature of the $s$--channel production. 
 
To measure the up--down asymmetry, we need a good  determination of the
reaction plane.  This is possible  for those events in which one of the
top quarks decays hadronically into jets.
One may be concerned about the imperfect angular resolution of hadronic
jets. We argue that even if the angles of jets are ambiguous at the level
of 10 degrees or so, as long as some sort of orientation of the reaction plane
can be defined, the asymmetry will not be smeared away by more than
one order of magnitude.  For example, one can simply discard events that 
the definition of up or down is made ambiguous by this smearing.
The details of such a smearing effect will strongly depend on future detectors.

 
\noindent
{V. \bf Higgs Model.}
 
Among various mechanisms for the CP violation, the one that may
manifest itself most easily is the neutral Higgs mediated CP violation.
Since the neutral Higgs couplings are typically proportional to the
quark mass, the large mass of the top quark naturally gives large
couplings to the neutral Higgs bosons. CP non--conservation occurs in
the complex Yukawa coupling,
\begin{equation} {\cal L}_{CPX}=-(m_t/v) \bar t (AP_L+A^*P_R) t H
   +  (m_Z^2/v)B H  Z^\nu Z_\nu
\;.
\end{equation}
Here $v=(\sqrt{2} G_F)^{-{1\over 2}} \simeq 246 \hbox{ GeV}$.
The complex coefficient $A$ is a combination of model-dependent mixing angles.
Simultaneous presence of both the real part $A_R=\rea A$ and the
imaginary part $A_I=\ima  A$ guarantees CP asymmetry. 
For example, in the low energy regime, it can give rise to the
electric dipole moment of elementary particles
\cite{ref:Weinberg,ref:Barr}.  Here we will derive the CP violating
form factors at high energy.
They are induced at the one--loop level as shown in the Fig.~3. First,
we calculate the absorptive parts according to the Cutkosky rules.
The leading contribution to Im$c_d^\gamma$ comes from the rescattering
of the top quark pair through the Higgs--boson exchange.
Both CP violation and the final state
effect are produced by the same one--loop graphs.
\begin{equation}
\ima  c_d^\gamma = c_v^\gamma \Bigl({m_t\over v}\Bigr)^2
                           {A_RA_I t^2\over 2\pi\beta}
              \Bigl(1-{h^2\over \beta^2} \log(1+{\beta^2\over h^2})\Bigr)
\;.
\end{equation}
The dimensionless variables are defined by,
$t=m_t/\sqrt{s}$, $z=m_Z/\sqrt{s}$, $\beta^2=1-4t^2$ as before,
as well as $h=m_H/\sqrt{s}$.
For Im$c_d^Z$, there is a similar contribution. In addition,
there could be a contribution due to the $ZH$ intermediate state, Fig~1b,
provided the kinematics is allowed.
\begin{equation}
\ima  c_d^Z= {c_v^Z \over c_v^\gamma}\ima  c_d^\gamma
            -{ \alpha A_IBc_v^Z t^2 \over 2(1-x_W)x_W\beta^2 }
           [\beta_Z+(2t^2+2t^2h^2-2t^2z^2-h^2)L]
\;.
\label{eq:zedm}
\end{equation}
Here $\beta_Z^2=1+h^4+z^4-2z^2-2h^2-2h^2z^2$, and
the logarithmic factor
\begin{equation}
 L={1\over \beta}\log{1-z^2-h^2 - \beta\beta_Z
                      \over
                       1-z^2-h^2 + \beta\beta_Z} \; .
\end{equation}
Our expression in Eq.(\ref{eq:zedm}) agrees with
that in Ref.\cite{ref:Bernreuther}.
Note that, at the threshold ($\beta =0$), $\ima c_d^\gamma$ vanishes
but $\ima c_d^Z$ has a value
$$
\ima  c_d^Z=
            -{ \alpha A_IBc_v^Z\beta_Z \over 8(1-x_W)x_W}
           \Bigl(  { 1-z^2+h^2 \over 1-z^2-h^2}
                 - {\beta_Z^2\over 3(1-z^2-h^2)^2} \Bigr)\;. $$
In case $m_Z+m_H<\sqrt{s}<2m_t$,
we still have the absorptive part $\ima c_d^Z$,
which is simply given by replacing the logarithm factor $L$ in
Eq.~(\ref{eq:zedm}) by its continuation,
\begin{equation}
 L=-{2\over\sqrt{-\beta^2}}\arctan{\beta_Z\sqrt{-\beta^2}
                                     \over 1-z^2-h^2} \;.
\end{equation}
The dispersive parts are obtained by the dispersion relation,
\begin{equation}
\rea c_d^j(s)={1\over\pi}P\int^\infty_{s_0}
                  {\ima c_d^j(s') \over s'-s} ds' \;.
\end{equation}
The symbol $P$ denotes the principal value of the singular integral.
Note that the absorptive part is evaluated at $s'$ above the threshold $s_0$,
which is $4m_t^2$ or $(m_Z+m_H)^2$ for $t\bar t$ or $ZH$
intermediate states respectively. It is understood that the
dimensionless variables, $\beta$, $\beta_Z$, $t^2$, $h^2$ and $z^2$, are
defined with respect to $s'$.
\centerline{\psfig{figure=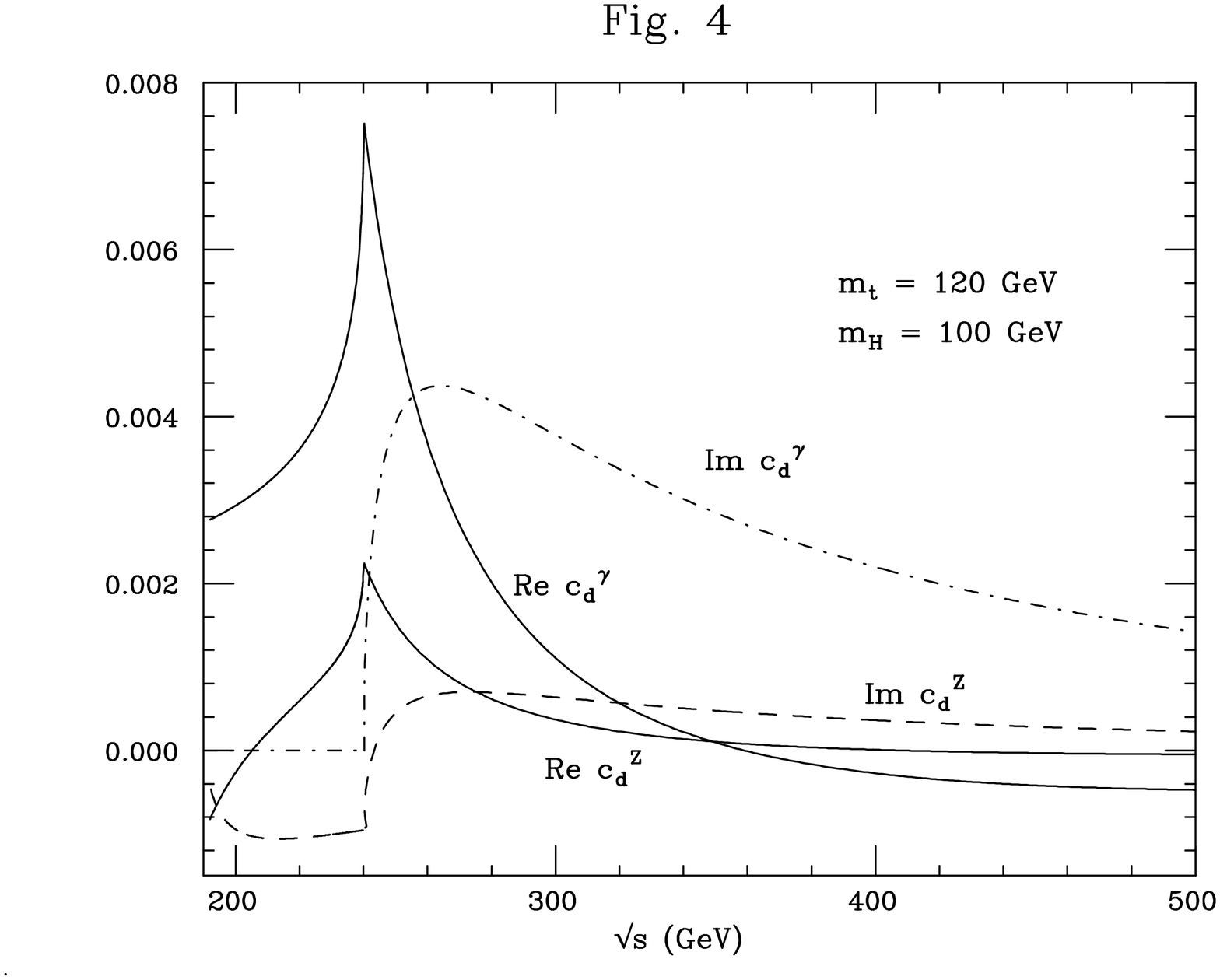,height=4in,width=6.5in,angle=90}}
Fig.~4 shows typical sizes of $c_d^j$ for various cases.
It is of order of $10^{-2}$ to $10^{-3}$ in gerneral.
We must keep in mind that we only show the Higgs boson contribution
in the perturbative regime. In scenarios that the Higgs bosons
interact strongly, the CP violating effect in the colliders could
be much larger.

\noindent
{\bf VI. Conclusion.}
 
We have shown that the energy asymmetry or the up--down
asymmetry of the prompt lepton events are sensitive
to the CP  violation in the top pair production in $e^+e^-$ annihilation.
These two asymmetries are complementary to each other for measuring the
absorptive and the dispersive parts of the CP violating form factors
$c_d^{\gamma,Z}$.
%
%
%

\centerline{\psfig{figure=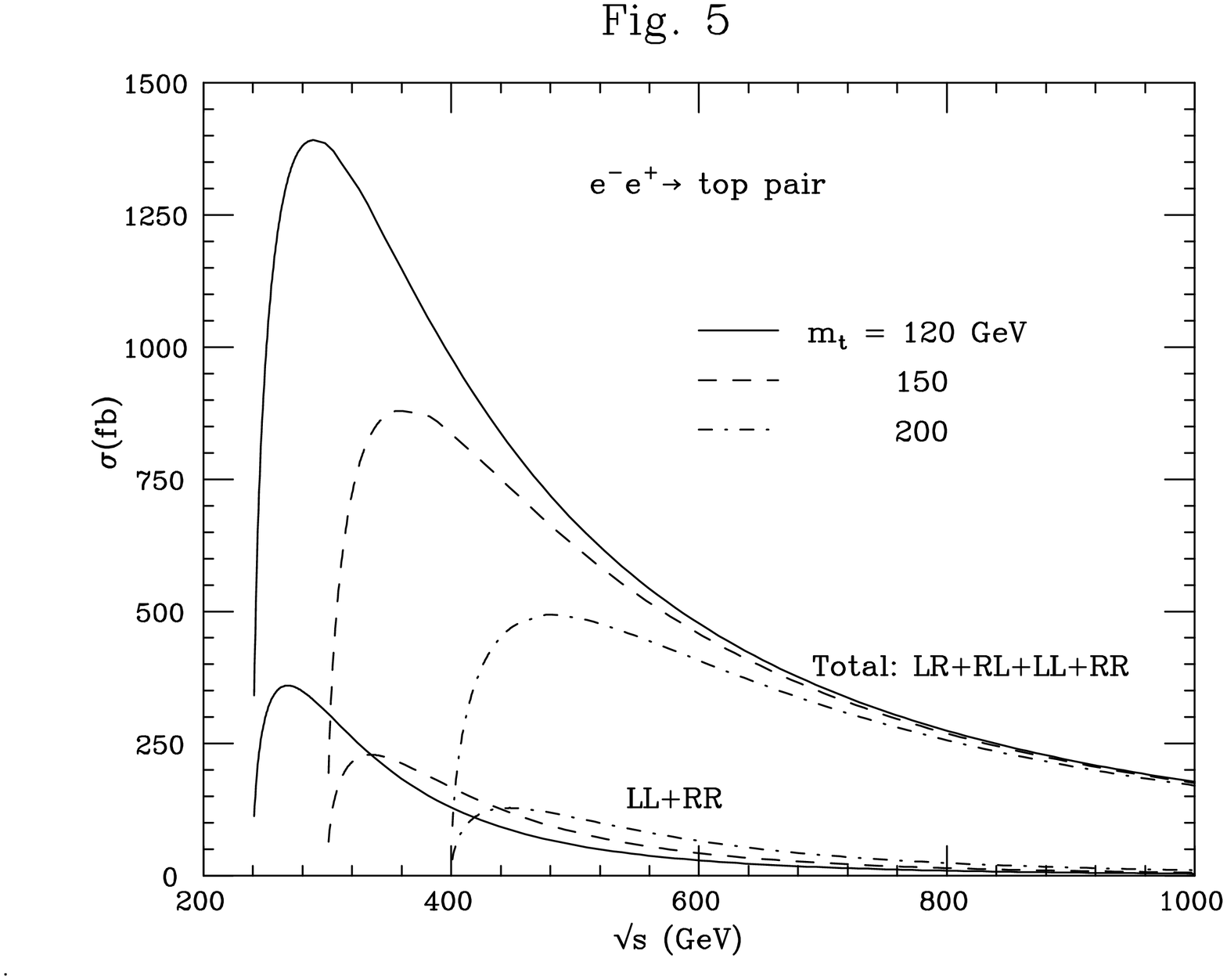,height=4in,width=6.5in,angle=90}}

The cross section for $e^-e^+\rightarrow t\bar t$ at $\sqrt{s}=400$ GeV,
for $m_t=170$ GeV, is $\sim 1400$~fb, among which about a quarter is due
to the channels $t_L\bar t_L$ and $t_R\bar t_R$ (see Fig.~5). At an
optimal luminosity of $10^{33}$ cm$^{-2}\cdot$s$^{-1}$, one year run of
the futuristic NLC\cite{ref:NLC} will produce 20,000 prompt semi--leptonic
events $b\bar b \ell^\pm X$. Note that we do not require both $t$ and
$\bar t$ to decay semileptonically. Therefore, only one power of the
branching fraction 2/9 is involved. Such sample is good enough to
analyze the CP violaing form factor $c_d$ at a level of ten percent. 
Although predictions from Higgs models is still smaller than this level
of sensitivity, the measurement could provide directly the important, 
model-independent information about the form factors of the top
quark interactions.
For a heavier top quark, $\sqrt{s}$ has to be raised above the
corresponding threshold, but not so far away as to incur
the $1/s$ suppression in event rates.
 
D.C. wishes to thank the Theory Group at the Institute of Physics at
Academia Sinica in Taipei, Taiwan and Department of Physics and
Astronomy at University of Hawaii at Manoa for hospitality while this
work was in progress.  This work is supported by grants from Department
of Energy and from National Science Council of Republic of China.

\section*{Figures}
\begin{enumerate}
\item The energy distributions of prompt leptons, for
the case that $m_t=170$ GeV, $\sqrt{s}=400$ GeV.
Case (a) for $N^{-1}[dN/dx(\ell^+)+dN/dx(\ell^-)]$,
and case (b) for $[dN/dx(\ell^+)-dN/dx(\ell^-)]/[dN/dx(\ell^+)+dN/dx(\ell^-)]$
per unit of Im$c_d^\gamma$ (solid) or Im$c_d^Z$ (dashed).

\item  The weighted up--down asymmetry of prompt leptons, for
the case that $m_t=170$ GeV, versus $\sqrt{s}$,
per unit of Re$c_d^\gamma$ (solid) or Re$c_d^Z$ (dashed).

\item Feynman diagrams for the  process $e^+e^-\rightarrow t\bar t$.
The tree amplitude interferes with those one--loop amplitudes with
(a) the final state interactions due to the exchange of a Higgs boson, or
(b) the intermediate state of the $ZH$ bosons.

\item
$c_d^\gamma$ and $c_d^Z$
versus $\sqrt{s}$  for the case
that $m_H= 100$ GeV and $m_t=120$ GeV.
The parameters are chosen to be $A_I=A_R=B=1$.

\item The total cross sections are shown in the upper three curves
for cases $m_t$=120 GeV (solid), 150 GeV (dashed), and 200 GeV 
(dash--dotted). The partial cross sections of the helicity configurations
$t_L\bar t_L$ and $t_R\bar t_R$ are shown in the lower three curves 
correspondingly.
\end{enumerate}
\begin{figure}[bht]
\begin{center}
\leavevmode
\epsfbox{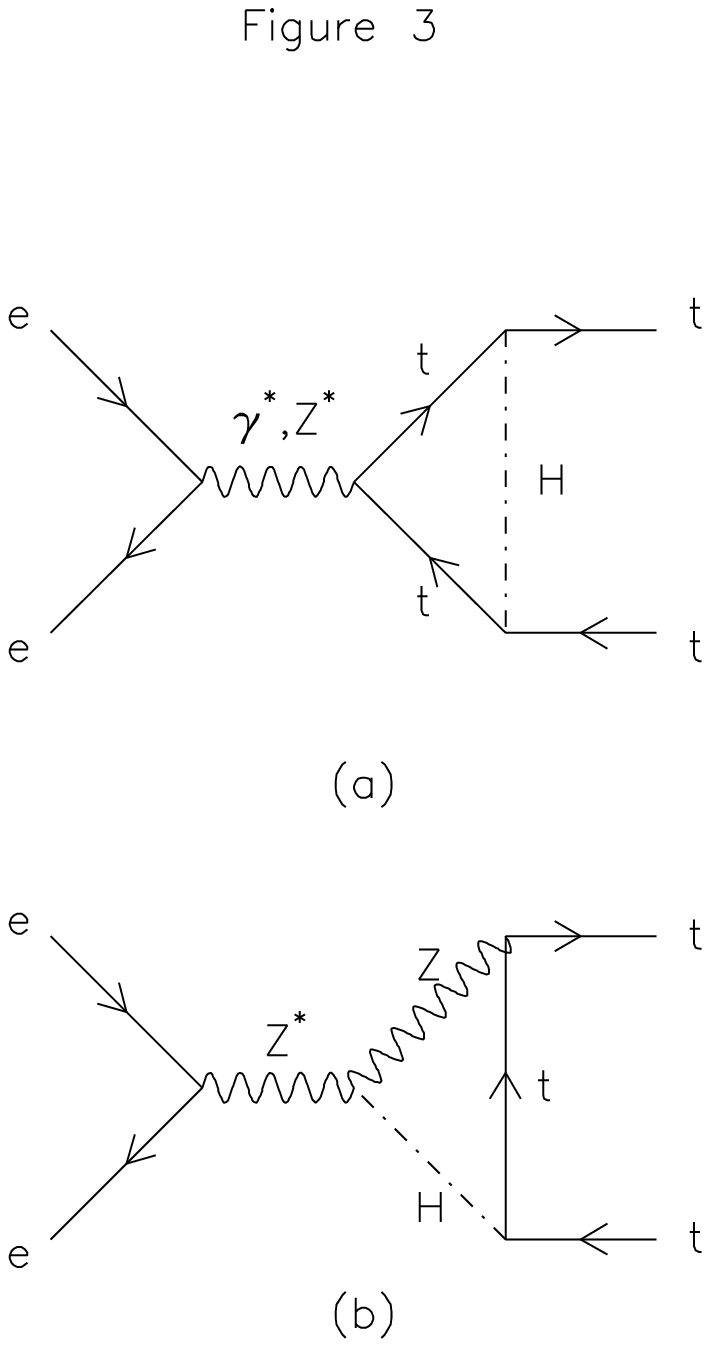}
\end{center}
\end{figure}

\begin{thebibliography}{99}
\bibitem{ref:phase}
Note that we use a different phase convention from Ref.\cite{ref:KLY}.
One can see sign flips between our Eq.(\ref{eq:amp}) and Eq.(6.6) in
Ref.\cite{ref:KLY}.
\end{thebibliography}
\end{document}